\documentclass {IEEEtran}

\IEEEoverridecommandlockouts
\usepackage{ifpdf}
\usepackage{cite}
\ifCLASSINFOpdf
   \usepackage[pdftex,final]{graphicx}
 \DeclareGraphicsExtensions{.eps}
\else
   \usepackage[dvips,final]{graphicx}
  \DeclareGraphicsExtensions{.eps}
\fi
\usepackage[cmex10]{amsmath}
\usepackage{algorithmic}
\usepackage{array}
\usepackage{mdwmath}
\usepackage{mdwtab}
\usepackage{eqparbox}
\usepackage{fancyheadings} 
\usepackage{url}
\usepackage{mdwmath}
\usepackage{mdwtab}
\usepackage{amsmath}

\usepackage{amssymb}
\usepackage{dsfont}
\begin{document}
\title{Multi-User Diversity with Optimal Power Allocation in Spectrum Sharing under Average Interference Power Constraint}
\author{\IEEEauthorblockN{Fotis~Foukalas$^{*}$, \textit{Member, IEEE,} Tamer~Khattab$^{*}$, \textit{Member, IEEE,} } \\
\IEEEauthorblockA{$^*$Electrical Engineering, Qatar University, Doha, Qatar\\}}

\maketitle

\begin{abstract}
In this paper, we investigate the performance of multi-user diversity (MUD) with optimal power allocation (OPA) in spectrum sharing (SS) under average interference power (AIP) constraint. In particular, OPA through average transmit power constraint in conjunction with the AIP constraint is assumed to maximize the ergodic secondary capacity. The solution of this problem requires the calculation of two Lagrange multipliers instead of one as obtained for the peak interference power (PIP) constraint and calculated using the well known water-filling algorithm. To this end, an algorithm based on bisection method is devised in order to calculate both Lagrange multipliers iteratively. Moreover, Rayleigh and Nakagami-$m$ fading channels with one and multiple primary users are considered to derive the required end-to-end SNR analysis. Numerical results are depicted to corroborate our performance analysis and compare it with the PIP case highlighting hence, the impact of the AIP constraint compared to the PIP constraint application.  
\end{abstract}


\IEEEpeerreviewmaketitle
\section{Introduction}
In a spectrum sharing (SS) system, the optimal power allocation (OPA) should satisfy the maximum allowable interference level at the primary system, additionally to the constraint on the average transmit power, and thereby to guarantee reliable operation for the primary users (PUs) as presented comprehensively in \cite{Musavian} and \cite{Xin}. This additional interference power constraint on the transmit power of the secondary transmitter (SU-Tx) can be assumed either as peak or averaged value, denoted as peak interference power constraint (PIP) and average interference power constraint (AIP) respectively. In case of the PIP application, the OPA results in the conventional method with one Lagrange multiplier, which can be iteratively calculated using the widely known water-filling algorithm; however, in case of the AIP application, the OPA yields to two Lagrange multipliers \cite{Xin}, and hence another type of algorithm is required to calculate both multipliers simultaneously. Moreover, an analysis is required in order to calculate the transmit power and the corresponding  performance of multi-user diversity (MUD). 

To this end, the contribution of this paper is two fold: a) the application of MUD with OPA in SS systems assuming AIP constraint by devising an algorithm based on bisection method in order to jointly calculate the two Lagrange multipliers and b) the end-to-end SNR analysis for the Rayleigh and Nakagami-$m$ fading channels with one PU and multiple PUs. The benefit of the AIP constraint over the PIP constraint is depicted through the numerical results. Notably, \cite{Zhang1} has highlighted the benefit of AIP over the PIP  in a single-user application scenario named as interference diversity. To our knowledge, such an investigation in a multi-user environment has not been provided so far, as we describe in a profound way in the following literature review. 

To be specific, Ban et al. in \cite{Ban} investigate the effects of MUD in an SS system applying PIP constraint for the protection of the primary receiver (PU-Rx). Based on this model, authors derive the ergodic capacity over Rayleigh fading channels for the high power SNR regime considering one PU-Rx and multiple PU-Rx. In \cite{Ekin}, Ekin et al. investigate the SS system proposed by Ban et al. in terms of hyper-fading communication channels for the secondary and primary links. They also derive the corresponding probability density function (PDF) and the cumulative density function (CDF) considering PIP constraint and peak transmit power. Li in \cite{Li1} and \cite{Li2} investigates opportunistic scheduling through MUD in SS systems with multiple SU-Rxs over Rayleigh fading channels and derives the outage and effective capacities considering PIP constraint as well as the uplink scenario in an SS system with multiple SU-Rxs deriving the achievable bit error rate and mean capacity using an outage capacity formulation. 

The rest of this paper is organized as follows. In Section II, the system model is presented. Section III gives the formulation of MUD with OPA in SS with AIP constraint and Section IV provides the end-to-end SNR over fading channels. Section V depicts and discusses the numerical results and finally, Section VI provides the conclusion of this work. 

\section{System Model} 

We consider a spectrum sharing (SS) system that consists of a secondary network with one secondary transmitter denoted as SU-Tx and $K$ secondary receivers denoted as SU-Rxs that utilizes a spectral band that is licensed to the primary system. The primary system is considered with $L$ multiple primary receivers denoted as PU-Rxs. The instantaneous channel power gains from the SU-Tx to the different SU-Rxs and PU-Rxs are denoted as $g_{s,i}$ and $g_{sp,j}$ respectively with $i\in[1,..,K]$ and $j\in[1,..,L]$. All channel gains are assumed to be independent and identically distributed (i.i.d.) exponential random variables with unit means in independent Rayleigh and Nakagami-$m$ fading channels and independent additive white Gaussian noise (AWGN) with random variables denoted as $n_{s,i}$ and $n_{sp,j}$ for the primary and secondary links respectively with mean zero and variance $N_0$ \cite{Andrea}.
	

The SU-Tx regulates its transmit power through the power control (PoC) mechanism that provides transmission with power constraints at both secondary and primary links in order to satisfy the requirements for transmission and protection at the SU-Rxs and the PU-Rxs simultaneously. We assume that the power constraints at the PoC of the SU-Tx are applied for keeping the transmit power budget at the secondary links under a predefined level, which is assumed as an average $P_{av}$ value, as well as for keeping the interference power at the primary links at a tolerable level, which again is assumed as an average $I_{av}$ value. We also assume that perfect channel state information (CSI) is available at the SU-Tx from the SU-Rxs and the PU-Rxs through a feedback channel \cite{Xin}. The SU-Tx provides multi-user diversity (MUD) in secondary network by which it is able to select for transmitting information among multiple SU-Rxs to the one with the best received SNR. The received SNR at the $i-th$ SU-Rx is given as $\gamma_{s,i}={P_t}g_{s,i}/{n_{s,i}^2}$, where the interference from the primary network is assumed negligible. 

\section{MUD with OPA in SS}
This section provides the analysis of MUD with OPA in SS. The OPA in SS when both average interference and transmit power constraints are considered is obtained as follows \cite{Musavian} \cite{Xin} : 
\begin{eqnarray}  \label{eq1}
P_t(g_{sp,j},g_{s,i}) = \left[ \frac{1}{\mu + \lambda g_{sp,j}} - \frac{n_{s,i}^2}{g_{s,i}}\right] 
\end{eqnarray}
where the Lagrange multipliers $\lambda$ and $\mu$ are related to the AIP constraint $I_{av}$ and the average transmit power constraint $P_{av}$ respectively. Notably, assuming PIP constraint, the OPA is related to the Lagrange multiplier $\lambda$ only \cite{Xin}, which is obtained using the well known water filling algorithm \cite{Andrea}. 

The calculation of Lagrange multipliers $\lambda$ and $\mu$ can be accomplished either separately or jointly using the following inequalities: 
\begin{eqnarray}  \label{eq2}
E[g_{sp,j} P_t(g_{sp,j},g_{s,i})]\leq I_{av}
\end{eqnarray}
and
\begin{eqnarray}  \label{eq3}
E[P_t(g_{sp,j},g_{s,i})]\leq P_{av}
\end{eqnarray}
Elaborating more on these calculations, we denote as $x=g_{sp,j}$, $y=g_{s,i}$, $z=\frac{g_{s,i}}{g_{sp,j}}$ and provide the following details in their analysis: 
\begin{eqnarray} \label{eq4}
& & E[g_{sp,j} P_t(g_{sp,j},g_{s,i})] \nonumber \\
&=& \int_{\mu}^{\mathcal{\infty}} \int_{\lambda}^{\mathcal{\infty}} x \left( \frac{1}{\lambda + \mu x} - \frac{1}{y} \right) f(x) f(y) dx dy \nonumber \\
&=& \int_{\mu}^{\mathcal{\infty}} \int_{\lambda}^{\mathcal{\infty}} \left( \frac{1}{\lambda + \mu x} - \frac{1}{z} \right) f(x) f(z) dx dz \leq I_{av}
\end{eqnarray}		
and 		
\begin{eqnarray} \label{eq5}
& & E[g_{sp,j} P_t(g_{sp,j},g_{s,i})] \nonumber \\
&=& \int_{\mu}^{\mathcal{\infty}} \int_{\lambda}^{\mathcal{\infty}} \left( \frac{1}{\lambda + \mu x} - \frac{1}{y} \right) f(x) f(y) dx dy \leq P_{av} . 
\end{eqnarray}
For the calculation of $I_{av}$ inequality, a new probability density function (PDF) denoted as $f(z)$ is required. This provided by \cite{Ghasemi} and \cite{Xin} for SS without incorporating MUD. Obviously, the end-to-end SNR analysis of PDF $f(z)$ for MUD, where  $z=\frac{g_{s,i}}{g_{sp,j}}$, has not been provided so far and this will be accomplished in this paper, in the following section, for Rayleigh and Nakagami-$m$ fading channels with one and multiple PU-Rxs. 

Regarding the calculation of $\lambda$ and $\mu$ Lagrange multipliers, it can be accomplished either separately or jointly. In order to calculate them jointly, we devise an algorithm which relies on bisection method as follows:
\begin{itemize}
\item \textbf{Given}  $\lambda \in [0,\bar{\lambda}]$ 
\item \textbf{Initialize}  $\lambda_{min} = 0$ and $\lambda_{max} = \bar{\lambda}$
\item \textbf{Repeat}
\begin{enumerate}
\item Set $\longleftarrow (1/2) (\lambda_{min}+\lambda_{max})$
\item Find the minimum $\mu$, $\mu > 0$, with which 
$E \left[ \frac{1}{\mu + \lambda g_{sp,j}} - \frac{n_{s,i}^2}{g_{s,i}}\right] \leq P_{av}$
\item Update $\mu$ by the bisection method: if $E \left[ \frac{1}{\mu + \lambda g_{sp,j}} - \frac{n_{s,i}^2}{g_{s,i}}\right] > I_{av}$, set $\lambda_{min} \longleftarrow \lambda$; otherwise, $\lambda_{max} \longleftarrow \lambda$. 
\end{enumerate}
\item \textbf{Until} $\lambda_{max}-\lambda_{min} \leq \epsilon$. 
\end{itemize} 
where $\epsilon$ is a small positive constant that controls the algorithm accuracy.

We assume now multi-user diversity (MUD), whereby the SU-Tx selects the SU-Rx with best channel quality among all  SU-Rxs. Thus, the received SNR of the selected SU-Rx $\gamma_{s,max}$ is obtained as follows \cite{Tse}:  
\begin{eqnarray} \label{eq6}
 \gamma_{s,max} = \max_{1\leq i \leq K} \ \gamma_{s,i}
\end{eqnarray}
with PDF given as follows:
\begin{eqnarray} \label{eq7}
 f_{\gamma_{s,max}}(x) = K f_{\gamma_{s,i}}(x)F_{\gamma_{s,i}}(x)^{K-1}
\end{eqnarray}
where $f_{\gamma_{s,i}}(x)$ and $F_{\gamma_{s,i}}(x)$ are the PDF and the CDF of the received SNR $\gamma_{s,i}$ at the $i-th$ SU-Rx respectively.  

The overall average achievable secondary capacity at the secondary system (i.e. SU-Tx to SU-Rx) when AIP constraint is considered, which results in \eqref{eq2}, is obtained as follows:
\begin{eqnarray} \label{eq8}
& & C_{s} = E[\log_2(1+\gamma_{s,max})] \nonumber \\
 &=& \int_{\mu}^{\mathcal{\infty}} \int_{\lambda}^{\mathcal{\infty}} {\ \textstyle{\log_2} (1+\gamma_{s,max}) f_{\gamma_{s,max}}(x) f_{\gamma_{s,max}}(z) dx dz} \nonumber \\
&=& \int_{\mu}^{\mathcal{\infty}} \int_{\lambda}^{\mathcal{\infty}} {\ \textstyle{\log_2} \left( \frac{g_{s,i}/n_{s,i}^2}{\lambda + \mu g_{sp,j}} \right) f_{\gamma_{s,max}}(x) f_{\gamma_{s,max}}(z) dx dz} \nonumber \\
\end{eqnarray}
and the corresponding outage probability as follows: 
\begin{eqnarray} \label{eq9}	
P_{out} &=& Pr \left\lbrace \gamma_{s,max}<\lambda \right\rbrace \nonumber \\
 &=& \int_{\mu}^{\mathcal{\infty}} \int_{\lambda}^{\mathcal{\infty}} f_{\gamma_{s,max}}(x) f_{\gamma_{s,max}}(z) dx dz .
\end{eqnarray}

\section{End-to-end SNR Analysis} 
In this section, we analyze the end-to-end SNR from the SU-Tx to SU-Rxs with AIP constraint to the link between the SU-Tx and PU-Rxs denoted as $z=\frac{g_{s,i}}{g_{sp,j}}$ $\forall i,j$. The particular analysis is obtained for Rayleigh and Nakagami-$m$ fading channels with one PU-Rx and multiple PU-Rxs. 

\subsection{Rayleigh with One PU-Rx} \label{Rayl2} 
Here, we assume that the channels gains $g_{s,i}$ and $g_{sp,j}$ are i.i.d. Rayleigh random variables  $\forall i,j$. For notational brevity, we will denote the term $g_{s,max}/g_{sp,j}$ as $g_s/g_{sp}$ and as before we will substitute $X=g_s/g_{sp}$ so that the PDF of the received SNR at the SU-Tx is obtained as follows: 
\begin{eqnarray} \label{eq10}
\nonumber
f(z) &=& \int_{0}^{\infty} \ h e^{-z h} e^{-h}  \,d h\\ 
&=& -\frac{e^{-(1+z)h}(1+h+z h)}{(1+z)^2}\mid_0^\infty = \frac{1}{(1+z)^2}
\end{eqnarray}
which is equal to the expression presented in \cite{Ghasemi}. The CDF of the PDF in  \eqref{eq10} is obtained as follows: 
\begin{eqnarray} \label{eq11} 
F(z) = 1 - \frac{1}{1+z} .
\end{eqnarray}
Substituting \eqref{eq10} and \eqref{eq11} into \eqref{eq7}, we can derive the PDF $f_{\gamma_{s,max}}(z)$ of the maximum received SNR $\gamma_{s,max}$ of the selected SU-Rx.

\subsection{Nakagami$-m$ with One PU-Rx} \label{Nakag} 
We now assume that the channels gains $g_{s,i}$ and $g_{sp,j}$ are i.i.d. Nakagami$-m$ random variables $\forall i,j$ and thus follow the following Nakagami$-m$ distribution for a specific channel gain:  
\begin{eqnarray}  \label{eq12} 
f(z) = \frac{m^m z^{(m-1)}}{\Gamma(m)}e^{(-m z)}, & &  z\geq 0
\end{eqnarray}
where $m$ represents the shape factor under which the ratio of the line-of-sight (LoS) to the multi-path component is realized. Assuming that both channels gains $g_{s,i}$ and $g_{sp,j}$ have instantaneously the same fading fluctuations i.e. $m_s=m_p=m$, the PDF of the term $z=g_s/g_{sp}$ is obtained as follows \cite{Ghasemi}:
\begin{eqnarray} \label{eq13} 
f(z) = \frac{z^{m-1}}{B(m,m)(z+1)^{2m}}, & & z \geq 0 .
\end{eqnarray}
Conducting thorough mathematical manipulation, the CDF of the PDF in \eqref{eq13} is obtained as follows: 
\begin{eqnarray} \label{eq14}
F_{g_s/g_p}(z) = \frac{1}{B(m,m)} \frac{z^m}{m} {_2}F_1(m,2m;1+m;-z)
\end{eqnarray}
where ${_2}F_1(a,b;c;y)$ is the Gauss hyper-geometric function which is a special function of the hyper-geometric series \cite{Functions}. 

\subsection{Rayleigh with Multiple PU-Rxs} 
We assume now that the spectrum is shared among $L$ PUs (i.e. PU-Rx) as mentioned in the system model. The PDF of the term $g_s/g_{sp}$ in Rayleigh fading channels with multiple PUs is given as follows \cite{Ghasemi}:  
\begin{eqnarray} \label{eq15}
f(z) = L\sum_{k=0}^{L-1} (-1)^k  \binom {L-1} {k} \frac{1}{(1+z+k)^2} .
\end{eqnarray}
Conducting thorough mathematical manipulation,  the CDF of the PDF in \eqref{eq15} is obtained as follows: 
\begin{eqnarray} \label{eq16}
F(z) = L \sum_{k=0}^{L-1} (-1)^k \binom {L-1} {k} \left(  \frac{1}{1+k} - \frac{1}{1+z+k} \right).
\end{eqnarray}

\subsection{Nakagami$-m$ with Multiple PUs} 
We assume the interference channel gains $g_{sp,j}$ between the SU-Tx and each $j=1,..,L$ PU-Rx are i.i.d. unit mean Nakagami random variables. We also assume that each transmit channel gain $g_{s,i}$ with $i=1,..,K$ is also independent of all the $g_{sp,j}$. We define now a variable for taking the maximum value of the interference channel denoted as $g_p$ that can be obtained as follows:
\begin{eqnarray} \label{eq17}
g_{sp} = \max_{j} g_{sp,j},& & j=1,\dot{..},L . 
\end{eqnarray}
Equation \eqref{eq17} represents the maximum value of all channel gains $g_{sp,j}$ between the SU-Tx and the PU-Rxs. The CDF of this value $g_{sp}$ is obtained as follows:
\begin{eqnarray} \label{eq18}
F_{g_{sp}}(x) = \prod_{j=1}^{L} F_{g_{sp,j}}(x)
\end{eqnarray}
where $F_{g_{sp,j}}(x)$ is the CDF of the Nakagami$-m$ fading distribution which can be derived from the integration of \eqref{eq12} as follows: 
\begin{eqnarray} \label{eq19}
\nonumber
F_{g_{sp,j}}(x) &=& \int_{0}^{x} \ \frac{m^m x^m-1}{\Gamma(m)}e^{-mx}  \,d x\\ 
&=& \frac{1}{\Gamma(m)} (1-\Gamma(m,mx)),  \forall j  .
\end{eqnarray} 
Therefore, substituting \eqref{eq19} into \eqref{eq18} the following is applied for the CDF of $g_{sp}$:
\begin{eqnarray} \label{eq20}
\nonumber
F_{g_{sp}}(x) &=& \prod_{j=1}^{L} \frac{1}{\Gamma(m)}(1-\Gamma(m,mx))\\
&=& \frac{1}{\Gamma(m)}(1-\Gamma(m,mx))^L .
\end{eqnarray}
We can now obtain the PDF of $g_{sp}$ by differentiating \eqref{eq20} as follows:
\begin{eqnarray} \label{eq21}
\nonumber
& & f_{g_{sp}}(x) = \frac{d}{dx} F_{g_{sp}}(x)\\ 
&=& {m^m}{e^{-mx}}{x^{m-1}} \frac{L}{\Gamma(m)^L} (1-\Gamma(m,mx))^{L-1} . 
\end{eqnarray}
We now proceed with the formulation of the PDF of the proportional channel gain used in SS systems  i.e. $g_s/g_{sp}$. Following the same procedure as in \cite{Ghasemi}, the PDF of the $z=g_s/g_{sp}$ is defined  in general as follows:
\begin{eqnarray} \label{eq22}
f_{g_s/g_{sp}}(z) = \int_{0}^{z} \ f_{g_s}(\frac{zy}{1+z}) f_{g_{sp}}(\frac{y}{1+z})\frac{y}{(1+z)^2}  \,d y
\end{eqnarray}
where $f_{g_s}(.)$ is the PDF at the secondary link obtained by \eqref{eq13} for the Nakagami$-m$ distribution and $f_{g_{sp}}(.)$  is the PDF at the primary link with the maximum channel gain value as obtained in \eqref{eq22}. Thus,equation \eqref{eq22} becomes: 
\begin{eqnarray} \label{eq23}
\nonumber
& & f_{g_s/g_p}(z) = \int_{0}^{\infty} \frac{m^m}{\Gamma(m)} (\frac{z y}{1+z})^{m-1} e^{-m\frac{z y}{1+z}} L\frac{m^m}{\Gamma(m)}\\
\nonumber
& & (\frac{y}{1+z})^{m-1} e^{-m\frac{y}{1+z}} (\frac{1}{\Gamma(m)}(1-\Gamma(m,m\frac{y}{1+z})))^{L-1}\\ 
\nonumber
& & \frac{y}{(1+z)^2} \,d y = L\frac{m^{2m}}{\Gamma(m)^{L+1}}\frac{z^{m-1}}{(1+z)^{2m}}\\
\nonumber
& & \int_{0}^{\infty} y^{2m-1}e^{-my}(1-\Gamma(m,m\frac{y}{1+z}))^{L-1} \,d y\\
\nonumber
&=& L\frac{m^{2m}}{\Gamma(m)^{L+1}}\frac{z^{m-1}}{(1+z)^{2m}} \int_{0}^{\infty} y^{2m-1} e^{-my}\\
\nonumber
& & \sum_{k=0}^{L-1} (-1)^k  \binom{L-1}{k} \Gamma(m,m\frac{k y}{1+z}) \,d y\\
\nonumber
&=& L\frac{m^{2m}}{\Gamma(m)^{L+1}}\frac{z^{m-1}}{(1+z)^{2m}} \sum_{k=0}^{L-1} (-1)^k  \binom{L-1}{k}\\
& & \int_{0}^{\infty} y^{2m-1}e^{-my} \Gamma(m,m\frac{ky}{1+z}) \,d y . 
\end{eqnarray} 
The integral in \eqref{eq23} is solved by using the equation (6.455.1) of the tables in \cite{Table}, which results in:                      
\begin{eqnarray} \label{eq24}
\nonumber
& & f_{g_s/g_p}(z) = L \frac{\Gamma(3m)}{\Gamma(m)^{L+1}} \frac{z^{m-1}}{2m} \sum_{k=0}^{L-1} (-1)^k \binom {L-1}{k} \\
& & \frac{k^m}{(z+k+1)^{3m}} {_2} F_1 (1,3m;2m+1;\frac{z+1}{z+k+1}) . 
\end{eqnarray}
In order to obtain the corresponding CDF for the Nakagami$-m$ distribution, we will do it numerically $\forall m$, since exact solution can not be easily derived. We give now an example of integrating the \eqref{eq24} substituting $m=2$ as follows:
\begin{eqnarray} \label{eq25}
\nonumber
& & F_{g_s/g_p}(z) = \int_{0}^{z}L \frac{\Gamma(6)}{\Gamma(2)^{L+1}} \frac{z}{4} \sum_{k=0}^{L-1} (-1)^k  \binom{L-1}{k}\\ 
\nonumber
&& \frac{k^2}{(z+k+1)^6} {_2} F_1(1,6;5;\frac{z+1}{z+k+1}) \,d z\\
\nonumber
& & = \int_{0}^{z}L \frac{\Gamma(6)}{\Gamma(2)^{L+1}} \frac{z}{4} \sum_{k=0}^{L-1} (-1)^k  \binom{L-1}{k}\\ 
\nonumber
&& \frac{k^2}{(z+k+1)^6} \frac{(1+k+z)(1+5k+z)}{5k^2} \,d z\\
\nonumber
& & = \frac{L\Gamma(6)}{\Gamma(2)^{L+1}20} \sum_{k=0}^{L-1} (-1)^k \binom{L-1}{k} (\frac{1+3k^2+4k}{6(1+k)^4}\\
& & -\frac{1+3k^2+4z+3z^2+4k(1+3z)}{6(1+k+z)^4}) . 
\end{eqnarray}
Notably, the use of a one-to-one mapping between the Ricean factor and the Nakagami fading parameter allows also Ricean channels to be well approximated by Nakagami$-m$ channels where the relation between them for $m=2$ gives a Ricean factor of 2.4312.  

\section{Numerical Results} \label{Result}
In this section, we provide the numerical results derived using the bisection-based algorithm and using the derived end-to-end SNR analysis assuming average transmit and interference power constraints, i.e. $P_{av}$ and $I_{av}$ respectively, over fading channels. 

Fig.2 depicts the average ergodic secondary capacity $C_s$ versus the AIP constraint $I_{av}$ with different numbers of SU-Rxs $K$ and PU-Rxs $L$ in the case of Rayleigh fading. We assume average transmit power $P_{av}=5dB$. Obviously, an increase in the number of SU-Rxs results in an increase in the capacity and on the other hand an increase in the number of PU-Rxs results in a decrease in capacity. This was expected since the probability that the SU-Tx will find an SU-Rx with the best SNR condition using MUD technique increases and the probability that the SU-Tx will find a PU-Rx in which the maximum interference level will be reached increases, leading to capacity increase and decrease respectively. A saturation on the performance is appeared due to the average transmit power $P_{av}$ constraint. Fig.3 depicts the corresponding outage probability $P_{out}$ of the implementation scenarios presented in Fig.2. In contrast with the capacity, the outage probability decreases when the number of SU-Rxs increases and increase when the PU-Rxs increases. 

\begin{figure}
  \centering
  \includegraphics[width=95mm,height=70mm]{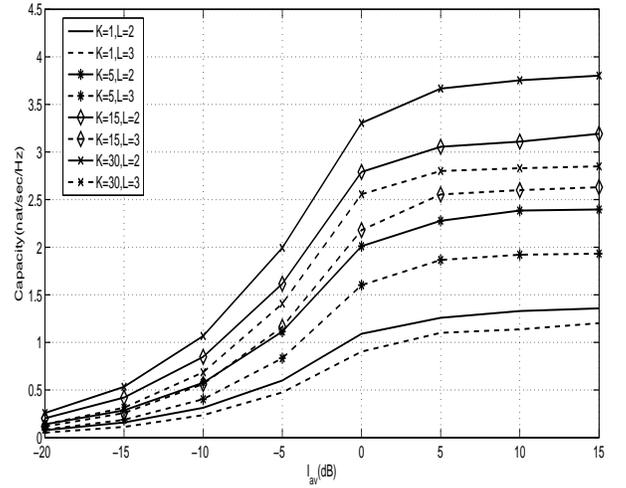}\\
  \caption{Capacity $C_s$ vs. average interference power constraint $I_{av}$ and $P_{av}=5dB$ with different numbers of SU-Rxs $K$ and PU-Rxs $L$.}
  \label{fig:cap1}
\end{figure}  

\begin{figure}
  \centering
  \includegraphics[width=95mm,height=70mm]{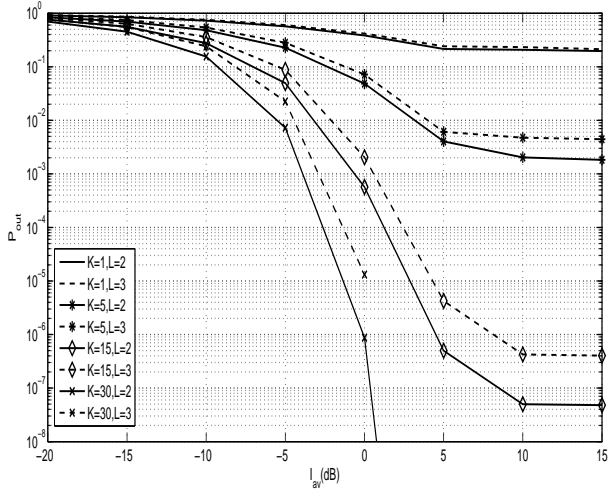}\\
  \caption{Outage probability $P_{out}$ vs. average interference power constraint $I_{av}$.}
  \label{fig:Pout1}
\end{figure}

Fig.4 depicts the ergodic secondary capacity $C_s$ vs. average transmit power $P_{av}$ assuming AIP constraint $I_{av}$ using the bisection-based algorithm and additionally depicts the results obtained using the PIP constraint $I_{pk}$, which follows conventional fading channels' analysis, i.e. the known Rayleigh and Nakagami-$m$ PDF and CDF. To be specific, we depict the results for $I_{pk}=5dB$ and $I_{av}=5dB$, for $K=5$ SU-Rxs and $L=1$ and $L=2$ PU-Rxs over $m=1$, i.e. Rayleigh, and $m=2$ fading channels. The outcome is that the impact of AIP vs. PIP increases when the number of PU-Rxs $L$ increases too. Therefore, as long as we have a high number of PU-Rxs, the benefit of AIP constraint is more significant due to a higher number of PU-Rxs, and thereby lower probability, which can satisfy the maximum AIP constraint in a long-term scenario. Finally, Fig.5 depicts the corresponding outage probability assuming the same settings for the MUD with OPA in SS. 

\begin{figure}
  \centering
  \includegraphics[width=95mm,height=70mm]{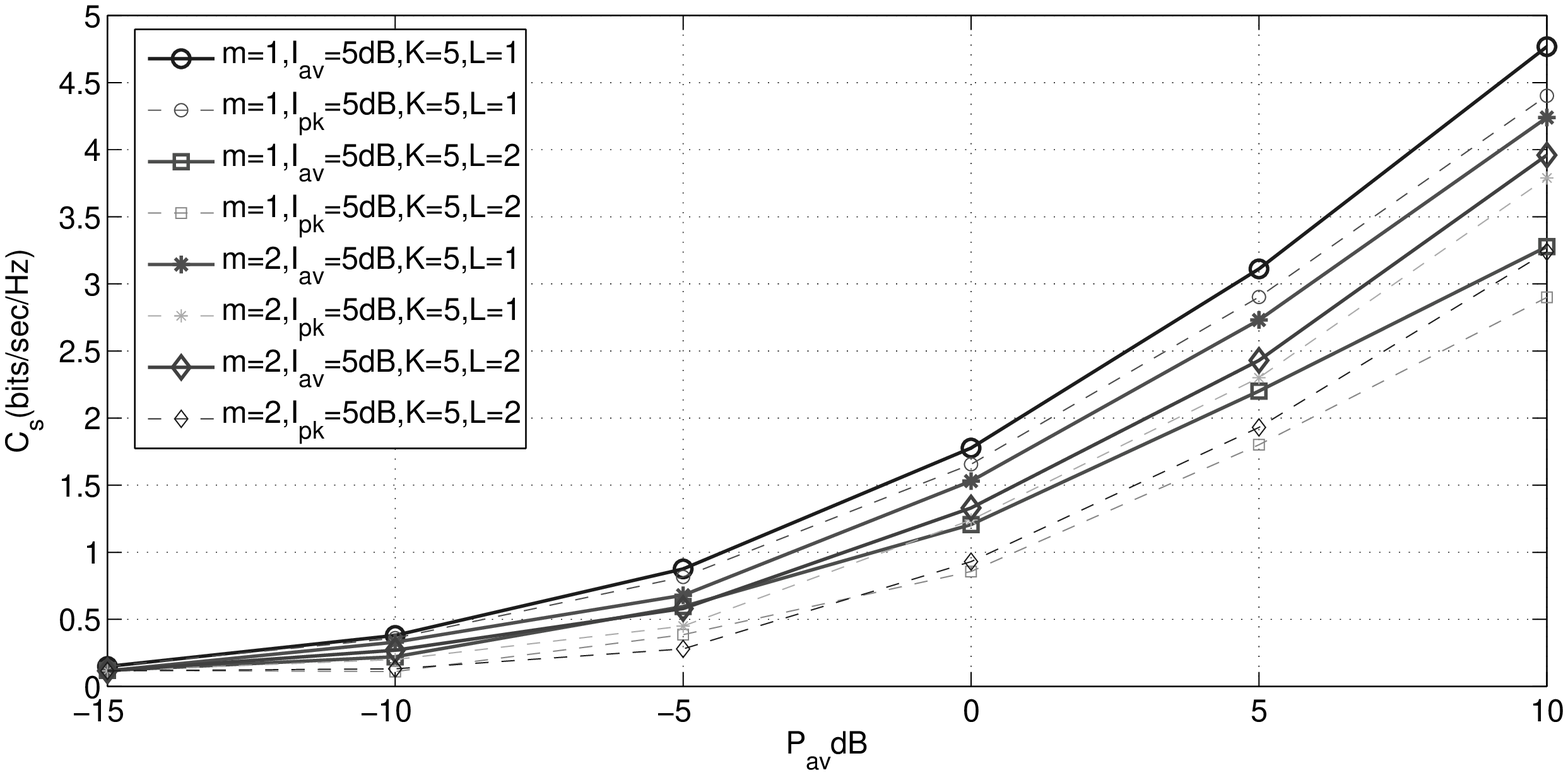}\\
  \caption{Capacity $C_s$ vs. average transmit power constraint $P_{av}$ for $K=5$ SU-Rxs, $L=1$ and $L=2$ PU-Rxs assuming $I_{pk}=5dB$ and $I_{av}=5dB$ with $m=1$ and $m=2$ Nakagami-$m$ fading channels.}
  \label{fig:cap2}
\end{figure}  

\begin{figure}
  \centering
  \includegraphics[width=95mm,height=70mm]{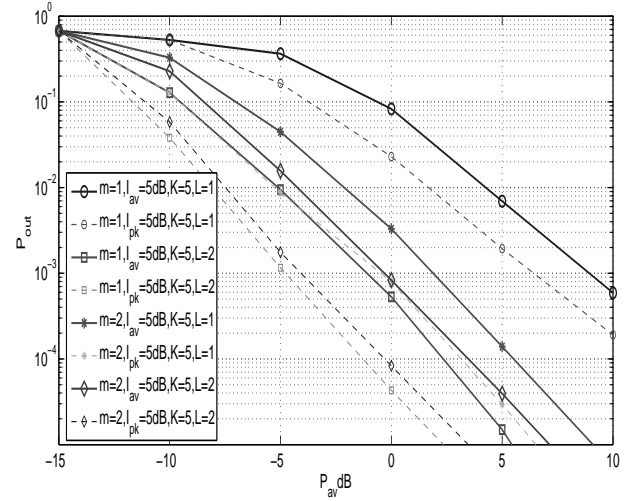}\\
  \caption{Outage probability $P_{out}$ vs. average transmit power constraint $P_{av}$ for $K=5$ SU-Rxs, $L=1$ and $L=2$ PU-Rxs assuming $I_{pk}=5dB$ and $I_{av}=5dB$ with $m=1$ and $m=2$ Nakagami-$m$ fading channels.}
  \label{fig:cap2}
\end{figure}  

 \section{Conclusion} \label{Result}
 In this paper, we analyzed the MUD with average interference and transmit power constraints, when multiple SU-Rxs and multiple PU-Rxs share the same channel, i.e. spectrum sharing cognitive radio system, over fading channels. Given the particular type of optimal power allocation of spectrum sharing when average interference and transmit power constraints are applied, we first devise an algorithm based on bisection method and second provide the end-to-end SNR analysis when MUD is incorporated. Rayleigh and Nakagami$-m$ fading channels are considered deriving the corresponding PDFs and CDFs. By obtaining numerical results, we concluded that the increase of PU-Rxs number can enhance the performance using AIP constraint that is more practically useful in case of tight interference power constraints, where more long-term chances to satisfy the maximum AIP in an average way for all PU-Rxs can be found and thereby to increase the capacity.

\end{document}